\documentclass[aps,preprint]{revtex4}
\begin{document}
\title{\Large \bf Five Dimensional Rotating Black Hole in a Uniform
Magnetic Field: The Gyromagnetic Ratio}
\author{\large A. N. Aliev}
\affiliation{ Feza G\"ursey Institute, P.K. 6  \c Cengelk\" oy,
81220 Istanbul, Turkey}
\author{\large Valeri {\bf P.} Frolov}
\affiliation{ Theoretical Physics Institute, Department of Physics,
University of Alberta, Edmonton, Canada T6G 2J1}
\date{\today}

\begin{abstract} In four dimensional general relativity, the fact
that a Killing vector in a vacuum spacetime serves as a vector
potential for a test Maxwell field provides one with an elegant way
of describing the behaviour of electromagnetic fields near a rotating
Kerr black hole immersed in a uniform magnetic field. We use a
similar approach to examine the case of a five dimensional rotating
black hole placed in a uniform magnetic field of configuration with
bi-azimuthal symmetry, that is aligned with the angular momenta of
the Myers-Perry spacetime. Assuming that the black hole may also
possess a small electric charge we construct the 5-vector potential
of the electromagnetic field in the Myers-Perry metric using its
three commuting Killing vector fields. We show that, like its four
dimensional counterparts, the five dimensional Myers-Perry black hole
rotating in a uniform magnetic field produces an inductive potential
difference between the event horizon and an infinitely distant
surface. This potential difference is determined by a superposition
of two independent Coulomb fields consistent with the two angular
momenta of the black hole and two nonvanishing components of the
magnetic field. We also show that a weakly charged rotating black
hole in five dimensions possesses two independent magnetic dipole
moments specified in terms of its electric charge, mass, and angular
momentum parameters. We prove that a five dimensional weakly charged
Myers-Perry black hole must have the value of the gyromagnetic ratio
$g=3$.

\end{abstract}
\maketitle

\section{Introduction}
Black holes originally predicted in four dimensional general relativity
have subsequently become an inseparable part of all higher dimensional
gravity theories. The fundamental features of black holes
in four dimensions, such as the equilibrium and uniqueness properties,
quantum properties following from Hawking's effect
of evaporation of microscopic black holes have revealed
an intimate connection between spacetime geometry, quantum field theory
and thermodynamics \cite{Israel, Carter, hawking1, hawking2}
(see Refs.\cite{fn, heusler, wald1} for comprehensive reviews).
Certainly, these properties of black holes
might have played a crucial role in the analysis of dynamics of
higher dimensional gravity theories, as well as in the compactification
process. For instance, to test the novel predictions of superstring
theory which, as believed, provides a consistent quantum theory of gravity
in higher dimensions \cite{hw}, microscopic black holes may serve
as good theoretical laboratories. Therefore over the last years
higher dimensional black holes have been widely considered as
very interesting objects to be investigated in detail.
Many interesting black hole solutions in various
higher-dimensional theories can be found in \cite{youm}.

On the other hand, the interest in higher dimensional black holes
has got a new impetus after the advent of brane-world gravity
theories \cite{ADD, AADD, RS1, RS2} (see also Refs. \cite{gabad, rubakov}).
The brane-world theories are built
on the basic idea that the geometry of our physical Universe is
indeed a "3-brane" - (3+1) dimensional hypersurface, embedded in fundamental
higher dimensional space. The size of the extra spatial dimensions
may be much larger than the convensional Planckian length
($\sim 10^{-33} cm $). Thus, in contrast to the original Kaluza-Klein's
scenario, in the brane-world models the
extra dimensions are supposed to manifest themselves as physical ones.
One of the dramatic consequences of these models is that
the fundamental scale of quantum gravity might become as low as
the weak interaction scale (of the
order of TeV). This, in turn, raises the problem of TeV-size black holes
and the exciting signature of such mini black holes is that they
can be directly probed in cosmic ray experiments or at future high
energy colliders \cite{dl,gt}. It has also been argued that one can
describe these black holes by the classical solutions of higher
dimensional vacuum Einstein equations provided that
the radius of the event horizon is much smaller than the size-scale
of the extra dimensions. In light of all this, it is obvious that further
knowledge of the special properties of black hole solutions in
higher dimensional vacuum gravity is of great importance.

The first black hole solution to the higher dimensional Einstein
equations is the static and hyperspherically symmetric
Schwarzschild-Tangherlini solution \cite{tang}, which has been
found a long time  ago within the Kaluza-Klein programme of
extension of four dimensional general relativity. Like, in four
dimensions, one could expect the generalization of the static
hyperspherical black hole solution to include rotational dynamics.
In 1986 Myers and Perry discovered the exact solution of
Einstein's equations describing such rotating black holes
\cite{mp}. It is important that the Myers-Perry solution is supposed
to be the most relevant to describe the "laboratory" black holes
in high-energy experiments \cite{gt}. In this context some essential
features of the Myers-Perry solution in five dimensions, such as
the existence of a Killing tensor and the separability of variables
in the Hamilton-Jacobi equations of motion, as well as quantum
radiation from a five dimensional black hole were explored in
\cite{fs1,fs2}.

However, the Myers-Perry solution is not unique
in five dimensions, unlike its four dimensional counterpart, the Kerr
solution. Emparan and Reall \cite{er} found a rotating
black ring solution in five dimensios with the horizon of toplogy
of $\; S^2 \times S^1\; $, which could have the same
mass and spin as the Myers-Perry solution. As for the static case,
the authors of \cite{gibbons1, gibbons2} have proved that in this case,
remarkably, the uniqueness property survives. Another essential
problem is the stability of higher
dimensional rotating black hole solutions. Recently in paper \cite{em}
it has been argued that the Myers-Perry solution becomes unstable for
the case of an arbitrary large rotation parameter for a fixed mass.
It is clear that, in the general case,
the full analytic theory of perturbations of higher dimensional
rotating black holes is needed to resolve the stabilitiy problem.
In the static case the stability of higher dimensional black holes
was proved in \cite{hideo}.

In this paper we shall study further properties of a five
dimensional rotating black hole in the presence of an external
magnetic field, which is supposed to be uniform at infinity.
We shall consider the configuration of the magnetic field
aligned with the angular momenta of the black hole. In other words,
the magnetic field shares the bi-azimuthal symmetry of the
Myers-Perry spacetime and has only two nonvanishing components.
In order to construct the
corresponding solution of the Maxwell field equations
in the background of the Myers-Perry metric, we shall appeal to
the well known fact \cite{papa} that in a vacuum spacetime one can
obtain the solution for the Maxwell test field by using only
isometries of that spacetime. Earlier, this fact was used in
four dimensional general relativity
to construct the solution for the electromagnetic field around a
Kerr black hole immersed in a uniform magnetic field \cite{wald}.
In this analysis the temporal and axial Killing  vectors of the
Kerr spacetime were used as a vector potential for the Maxwell test
field. In particular, it has been found that the so-called Wald
effect takes place; the rotation of a black hole
in an asymptotically uniform magnetic field causes an inductive
potential difference between the event horizon and infinity, thereby
the black hole may acquire an inductive electric charge.

In four dimensions the rotation group is $\; SO(3)\;$ and there always
exists a rotation axis consistent with only one independent
Casimir invariant. However, in five dimensions the rotation group
is $\; SO(4)\;$, which posseses two independent Casimir
invariants. These two Casimir invariants, in turn, are associated
with two independent rotations of the system. In other words,
a rotating black hole in five dimensions may have two distinct
planes of rotation specified by appropriate azimuthal coordinates,
rather than an axis of rotation. In accordance with this,
the stationary and asymptotically flat Myers-Perry metric admits
three commuting Killing vector fields, which reflect
the time-translation invariance and bi-azimuthal symmetry of this metric
in five dimensions. We shall use these Killing vectors to construct
the 5-vector potential for a test electromagnetic field, when
the Myers-Perry black hole is placed in an asymptotically
uniform magnetic field of a configuration with bi-azimuthal
symmetry. We shall show that a rotation
in two distinct planes of a five dimensional black hole immersed
in a uniform magnetic field produces an inductive electric field
which is determined by the superposition of two independent
Coulomb parts consistent with the two angular momentum parameters and
two nonvanishing components of the magnetic field.

We shall also examine the case, when a five dimensional
black hole may have an electric charge small enough
that the spacetime is still well described
by the Myers-Perry solution. In this case the rotation of the black hole
must produce a dipole magnetic field. We shall establish that the
black hole, in fact, possesses two independent magnetic dipole
moments determined only by its charge, mass and angular momentum.

The paper is organized as follows. In Sec.II we recall some of
the properties of the Myers-Perry metric for a five dimensional
black hole. We describe the Killing isometries of the metric, its
mass parameter, and specific angular momenta. We also  give the
precise definition of the angular velocities for stationary observers in
the Myers-Perry metric and define locally non-rotating orthonormal
frames (LNRFs) and their dual basis forms. In Sec.III we begin with a
brief description of a uniform magnetic field in five dimensions and
construct the 5-vector potential for the Maxwell test field using the
Killing isometries of the Myers-Perry spacetime. Here we also
calculate the magnetic flux crossing a portion of the black hole
horizon. The dominant orthonormal components of electric and magnetic
fields in the asymptotic rest frame of a weakly charged Myers-Perry
black hole, as well as its magnetic dipole moments are calculated in
Sec.IV . Finally, in Sec.V we prove that a five dimensional weakly
charged Myers-Perry black hole must possess the value of the gyromagnetic
ratio $g=3$.

\section{Five Dimensional Rotating Black Hole}

\subsection{Myers-Perry metric}

The metric of a rotating black hole in five dimensions follows from
the general asymptotically flat solutions to  $\;(N+1)\;$ dimensional
vacuum gravity found by Myers and Perry \cite{mp}. In Boyer-Lindquist
type coordinates it takes the most simple form given by
\begin{eqnarray}
ds^2 & = & -dt^2 +\Sigma\,\left(\frac{r^2}{\Delta}\,dr^2 + d\theta^2 \right)
+(r^2 + a^2)\,\sin^2\theta \,d\phi^2
+(r^2 + b^2)\,\cos^2\theta \,d\psi^2 \nonumber \\
&  & +\,\frac{m}{\Sigma}\, \left(dt - a\, \sin^2\theta \,d\phi
- b\,\cos^2\theta \,d\psi \right)^2\,\,,
\label{metric}
\end{eqnarray}
where
\begin{equation}
\Sigma=r^2+a^2 \,\cos^2\theta + b^2 \,\sin^2 \theta, \;\;\;\;\;\;
\Delta= (r^2 + a^2)(r^2 + b^2) -m \, r^2 ,
\end{equation}
and $\,m \,$  is a parameter related to the physical
mass of the black hole, while the parameters $\,a \,$ and $\,b \,$
are associated with its two independent angular momenta. For the
metric determinant we have
\begin{equation}
\sqrt{-\,g}=r \,\Sigma\,\sin\theta \cos\theta
\label{determinant}
\end{equation}

The components of the inverse metric have the following forms:
\begin{eqnarray}
g^{tt}&=&-\left(1+ \frac{m}{\Sigma} + \frac{m\,r^2}{\Delta\,\Sigma}\right)\,,
~~~~~
g^{rr}=\frac{\Delta}{r^2\,\Sigma}\,,~~~~~g^{\theta\theta}=\frac{1}{\Sigma}\,,
~~~~~
g^{\phi \psi} =-\frac{m\,a\,b}{\Delta\,\Sigma}\,,
\nonumber \\[3mm]
g^{t\phi}& =&-\,\frac{m\,a\,(r^2+b^2)}{\Delta\,\Sigma}\,,~~~~~~~~
g^{\phi \phi} =\frac{1}{\Sigma}\,\left[\frac{1}{\sin^2\theta} +
\frac{(r^2+b^2)(b^2-a^2)- m\,b^2}{\Delta}\right]\,, \nonumber \\[3mm]
g^{t\psi}& =&-\,\frac{m\,b\,(r^2+a^2)}{\Delta\,\Sigma}\,,~~~~~~~~
g^{\psi \psi} =\frac{1}{\Sigma}\,\left[\frac{1}{\cos^2\theta} +
\frac{(r^2+a^2)(a^2-b^2)- m\,a^2}{\Delta}\right]\,,
\label{contra}
\end{eqnarray}

The event horizon of the black hole is a null surface determined by
the equation $\,g^{rr}=0\,$,  which implies that
\begin{eqnarray}
\Delta&= &(r^2 + a^2)(r^2 + b^2) -m \, r^2 =0 \,\,.
\label{nullsurf}
\end{eqnarray}
The largest root of this equation gives the radius of the black hole's
outer event horizon. We have
\begin{equation}
r_{h}^2 = \frac{1}{2}\,\left(m - a^2-b^2 +
\sqrt{(m - a^2- b^2)^2 - 4\,a^2\,b^2} \right)
\label{hradius}
\end{equation}
Notice that the horizon exists if and only if
\begin{equation}
a^2 + b^2 + 2 |a\,b| \,\leq m\,\,,
\label{extreme}
\end{equation}
so that the condition $ \, m=a^2 + b^2 + 2 |a\,b| \,$ or, equivalently,
$\,r_{h}^2= |a\,b|\,$  defines the extremal horizon of a
five dimensional black hole.

In the absence of the black hole $ (\,m=0 \,) $, the metric (\ref{metric})
reduces to the flat one written in oblate bi-polar coordinates. The latter
can be readily cast in the Minkowski form using the transformation
of coordinates \cite{fs1}
\begin{eqnarray}
x&=&\sqrt{r^2+a^2}\,\sin\theta\, \cos\tilde \phi ,~~~
y= \sqrt{r^2+a^2}\,\sin\theta\, \sin\tilde \phi \nonumber \\
z&= &\sqrt{r^2+b^2}\,\cos\theta\, \cos\tilde \psi , ~~~
w= \sqrt{r^2+b^2}\,\cos\theta\, \sin\tilde \psi
\label{cartesian}
\end{eqnarray}
where
$$ \tilde \phi= \phi + \tan^{-1} (a/r)\,~~~
\tilde \psi= \psi + \tan^{-1} (b/r) \; . $$
On the other hand, for $ \, a=b=0 \, $  we have the
Schwarzschild-Tangherlini static solution in spherical
bi-polar coordinates. It is clear that in general,
the metric (\ref{metric}) admits two orthogonal 2-planes of
rotation (x-y plane, z=w=0) and (z-w plane, x=y=0),
which are specified by the azimuthal
angles $\,\phi\,$ and $\,\psi\,$ respectively. These angles both range
from $\,0\,$ to $\,2\,\pi\,$, while $\,\theta\,$ being the angle between
the two orthogonal 2-planes varies in the interval
$ \,\left[0,\,\pi/2\,\right].$  As a consequence, the metric reveals
the following obvious invariance properties;
under the simultaneous inversion
of the time coordinate $\,t \rightarrow -\,t $ and the angles
$\,\phi \rightarrow -\,\phi $ \,,\,\, $\,\psi \rightarrow -\,\psi $ ,
and  under the transformation
\begin{equation}
a\,\leftrightarrow \,b ,~~~~
\phi\,\leftrightarrow \,\psi ,~~~~ \theta\,\leftrightarrow \,
\frac{\pi}{2}-\theta \,.
\label{symmetry}
\end{equation}

The bi-azimuthal symmetry properties of the five dimensional black hole
metric (\ref{metric}) along with its stationarity imply the existence
of the three commuting Killing vectors
\begin{equation}
{\bf \xi}_{(t)}= \partial / \partial t\,, ~~~~
{\bf \xi}_{(\phi)}= \partial / \partial \phi \, , ~~~~
{\bf \xi}_{(\psi)}= \partial / \partial \psi \,.
\label{killing}
\end{equation}
The various scalar products of these Killing vectors are expressed
through the corresponding metric components as follows
\begin{eqnarray}
{\bf \xi}_{(t)} \cdot {\bf \xi}_{(t)}&=& g_{tt}= -1+ \frac{m}{\Sigma}\,\,,
~~~~~~~~~
{\bf \xi}_{(\phi)} \cdot {\bf \xi}_{(\phi)}= g_{\phi\phi}=
(r^2+a^2 + \frac{m\,a^2}{\Sigma}\,\sin^2 \theta)\,\sin^2\theta\,,
\nonumber \\[3mm]
{\bf \xi}_{(t)} \cdot {\bf \xi}_{(\phi)}&=& g_{t\phi}=
-\,\frac{m\,a}{\Sigma}\, \sin^2 \theta\,,~~~~~~
{\bf \xi}_{(\psi)} \cdot {\bf \xi}_{(\psi)}= g_{\psi\psi}=
(r^2+b^2 + \frac{m\,b^2}{\Sigma}\,\cos^2 \theta)\,\cos^2\theta   \,,
\nonumber \\ [3mm]
{\bf \xi}_{(t)} \cdot {\bf \xi}_{(\psi)}&=& g_{t\psi}=
-\,\frac{m\,b}{\Sigma}\, \cos^2 \theta\,,~~~~~~
{\bf \xi}_{(\phi)} \cdot {\bf \xi}_{(\psi)}= g_{\phi\psi}=
\frac{m\,a\,b}{\Sigma}\, \sin^2 \theta\,\cos^2 \theta\,.
\label{kproduct}
\end{eqnarray}

The Killing vectors (\ref{killing}) can be used to give a
physical interpretation of the parameters $\,m\,$\,,
$\,a\,$ and $\,b\,$ involved in the metric (\ref{metric}).
Namely, following the analysis given in \cite{komar} one can obtain 
coordinate-independent definitions for these parameters. We have
the integrals
\begin{equation}
m=\frac{1}{4\,\pi^2}\,\oint \,\xi_{(t)}^{\mu\,;\,\nu}\,
d\,^3 \Sigma_{\mu\;\nu}
\label{mass}
\end{equation}
and
\begin{eqnarray}
j_{(a)} & =&a\,m=-\,\frac{1}{4\,\pi^2}\,\oint \,\xi_{(\phi)}^{\mu\,;\,\nu}\,
d\,^3 \Sigma_{\mu\;\nu}\,,~~~~~~~~
j_{(b)}  =b\,m =-\,\frac{1}{4\,\pi^2}\,\oint \,\xi_{(\psi)}^{\mu\,;\,\nu}\,
d\,^3 \Sigma_{\mu\;\nu}\,,
\label{momenta}
\end{eqnarray}
where the integrals are taken over the $3$-sphere at spatial infinity,
\begin{equation}
d\,^3 \Sigma_{\mu\;\nu}= \frac{1}{3!}\,\sqrt{-g}\,
\epsilon_{\mu\,\nu\,\alpha\,\beta\,
\gamma} \,d\,x^{\alpha} \wedge d\,x^{\beta} \wedge d\,x^{\gamma}\,\,,
\label{3sphere}
\end{equation}
the semicolon denotes covariant differentiation
and we have introduced the two specific angular momentum
parameters $\,j_{(a)}\,$ and $\,j_{(b)}\,$ associated with rotations
in the $\,\phi \,$ and $\,\psi \,$ directions respectively.
We note that with these definitions the relation between the
specific angular momentum
and the mass parameter looks exactly like the corresponding relation
$\,(J=a\,M) \,$ of four dimensional Kerr metric.

To show that the definitions given in (\ref{mass}) and (\ref{momenta})
do in fact correctly describe the mass and angular momenta parameters
we can calculate the integrands in the asymptotic
region $\,r \rightarrow \infty\,$. The dominant terms in the asymptotic
expansion have the form
\begin{eqnarray}
\xi_{(t)}^{t\,;\,r}& = & \frac{m}{r^3}\, +
\mathcal{O}\left(\frac{1}{r^5}\right)\,\,,
\nonumber \\[4mm]
\xi_{(\phi)}^{t\,;\,r}& = & -\,\frac{2\,a\,m\,\sin^2\theta}{r^3}\,
+ \mathcal{O}\left(\frac{1}{r^5}\right)\,\,,~~~~~~~
\xi_{(\psi)}^{t\,;\,r} = -\, \frac{2\,b\,m\,\cos^2\theta}{r^3}\,
+ \mathcal{O}\left(\frac{1}{r^5}\right)\,\,.
\label{mjj}
\end{eqnarray}
The substitution of these expressions into the formulas
(\ref{mass}) and (\ref{momenta}) verifies them. On the other hand,
the relation of the above parameters to the total mass $\,M\,$
and the total angular momenta $\,J_{(a)}\,$ and $\,J_{(b)}\,$
of the black hole can be established using the formulas
given in \cite{mp}. We obtain that
\begin{equation}
m=\frac{8}{3\,\pi}\,M\,\,,~~~~~j_{(a)}=\frac{4}{\pi}\,\,J_{(a)}\,\,,~~~~~
j_{(b)}=\frac{4}{\pi}\,\,J_{(b)}\,.
\label{physical}
\end{equation}
These relations confirm the interpretation of the parameters
$\,m\,$\,, $\,a\,$ and $\,b\,$  as being related to the physical
mass and angular momenta of the metric (\ref{metric}).

\subsection{Locally Nonrotating Observers}

To examine further properties of the five dimensional Myers-Perry
black hole, as well as physical processes near such a black hole, it
is useful to introduce a family of locally nonrotating
observers. In the four-dimensional case a locally nonrotating
observer has a vector of velocity orthogonal to the $t=$ const surface in
the Kerr geometry and its  angular momentum vanishes \cite{bard,mtw}.
We define a locally nonrotating observer in the Myers-Perry metric in
a similar manner. Let us write
\begin{equation}
u^{\mu}=u^{\mu}(r,\theta)=\alpha\, \left(
\xi_{(t)}^{\mu}+\Omega_{(a)}\xi_{(\phi)}^{\mu}
+\Omega_{(b)}\xi_{(\psi)}^{\mu}\right)\, ,
\end{equation}
where $\,u^{\mu}\,$ is a unit vector of 5-velocity of a locally
nonrotating observer, and
$\alpha$ is a normalization constant determined by the condition
$u^2=-1$. The orthogonality to the $t=$const surface implies
$u^{r}=u^{\theta}=0$ and
\begin{eqnarray}
g_{t \phi}\,u^{t} + g_{\phi\phi}\,u^{\phi} +
g_{\phi\psi}\,u^{\psi} & =& 0\,,
\nonumber \\ [3mm]
g_{t \psi}\,u^{t} +   g_{\psi\psi}\,u^{\psi}
+ g_{\psi\phi}\,u^{\phi} & =& 0 \,\,.
\label{eqs1}
\end{eqnarray}

The simultaneous solution of these equations
determines $u^{\mu}(r,\theta)$. Thus we obtain
\begin{eqnarray}
\Omega_{(a)} & =& \frac{u^{\phi}}{u^t}\; = \;
\frac{g_{t\psi}\,g_{\phi\psi}-g_{t\phi}\,
g_{\psi\psi}}{g_{\phi\phi}\,g_{\psi\psi}-{g_{\phi\psi}}^2}\,
=\,\frac{a\,m\,(r^2+b^2)}{\Delta\,\Sigma + m\,(r^2+a^2)(r^2+b^2)}
\,\,,\nonumber \\ [3mm]
\Omega_{(b)} & =& \frac{u^{\psi}}{u^t}\; = \;\frac{g_{t\phi}\,g_{\phi\psi}
-g_{t\psi}\,g_{\phi\phi}}{g_{\phi\phi}\,g_{\psi\psi}-{g_{\phi\psi}}^2}\,
=\, \frac{b\,m \,(r^2+a^2)}{\Delta\,\Sigma + m\,(r^2+a^2)(r^2+b^2)} \,\,.
\label{angvel}
\end{eqnarray}

In the case that either $\,b=0\,$, or $\,a=0\,$  instead of the above
expressions  we have either
\begin{eqnarray}
\Omega_{(a)}& = &
-\frac{g_{t\phi}}{g_{\phi\phi}}\,\,, ~~ \mbox{\bf or}  ~~~~~~~~
\Omega_{(b)} = -\frac{g_{t\psi}}{g_{\psi\psi}}
\label{angvel1}
\end{eqnarray}
which are closly reminiscent of the corresponding relation for
the Kerr black hole in four dimensions. At large distances
from the black hole, the relations (\ref{angvel}) can be written in the form
\begin{eqnarray}
\Omega_{(a)} & =&
\, \frac{j_{(a)}}{r^4} + \,\mathcal{O}\left(\frac{1}{r^6}\right)\,\,,
~~~~~~~  \Omega_{(b)}  =
\, \frac{j_{(b)}}{r^4} + \,\mathcal{O}\left(\frac{1}{r^6}\right)\,\,,
\label{angvel2}
\end{eqnarray}
which reveals the remarkable property of the Myers-Perry black hole, namely
{\it the dragging of inertial frames} in both $\,\phi\,$
and $\,\psi\,$ $2$-planes of rotation. Clearly, the effect
of "bi-dragging" disappears at spatial infinity. However, toward the
event horizon of the black hole it increases tending to its
constant value on the event horizon $\,(\Delta=0)\,$. From equations
(\ref{angvel}) we obtain
\begin{equation}
\Omega_{(a)\,h} = \frac{a}{r_{+}^2 +a^2}\,\,,~~~~~~~
\Omega_{(b)\,h} = \frac{b}{r_{+}^2 +b^2}\,\,.
\label{horizon}
\end{equation}
These quantities can be interpreted as angular velocities of the black hole
\cite{mp}. In order to show this, one needs to know the
isometry properties of the horizon geometry. Following the Hawking
approach in four dimensional
geometry of the Kerr black hole \cite{hawking1}, we suppose that
the isometry of the event horizon of a five dimensional black hole is
described by a Killing vector, which must be a linear combination of
the three Killing vectors given in (\ref{killing}). Thus, we can take it
in the form
\begin{equation}
\chi = {\bf \xi}_{(t)}+ \Omega_{(a)\,h}\,{\bf \xi}_{(\phi)}+
\Omega_{(b)\,h} \,{\bf \xi}_{(\psi)}\,\,.
\label{khorizon}
\end{equation}
One can easily verify that this vector becomes null at the surface
$\,\Delta=0 \,$, i.e. it is tangent to the null surface of the horizon.
This means that stationary observers moving in the $\,\phi\,$ and
$\,\psi\,$ $2$-planes of rotation become co-rotating along with the
horizon with the local angular velocities (\ref{horizon}).
In other words, the event horizon of the five dimensional Myers-Perry
black hole is, in fact, the Killing horizon determined
by the vector $\,\chi\,$.

From what we said above about the locally nonrotating observers it also
follows that with any such observer one can associate an orthonormal
frame. The form of the metric given in (\ref{metric})
enables us to choose the appropriate basis $1$-forms of these frames.
In particular, for a locally nonrotating observer moving in
the $\,\phi\,$ $2$-plane we can choose an orthonormal frame $\,(LNRF)\,$
with the following basis $1$-forms
\newpage
\begin{eqnarray}
\omega\,^{\hat{t}} & =& \left| g_{tt}- \frac{{g_{t\psi}}^2}{g_{\psi\psi}}
-\Omega_{(a)}^2\,\,\frac{g_{\phi\phi}\,g_{\psi\psi}-{g_{\phi\psi}}^2}
{g_{\psi\psi}} \right|^{1/2}\,d\,t \nonumber\\[3mm]
\omega\,^{\hat{\phi}} & =& \left(\frac{g_{\phi\phi}\,g_{\psi\psi}
-{g_{\phi\psi}}^2} {g_{\psi\psi}} \right)^{\,1/2} \,\left(d\,\phi-
\Omega_{(a)}\,d\,t\right)
\nonumber\\[3mm]
\omega\,^{\hat{\psi}} & =& (\,g_{\psi\psi}\,)\,^{1/2}\,
\left(d\,\psi +\frac{g_{t \psi}}{g_{\psi\psi}}\,d\,t+
\frac{g_{\phi \psi}}{g_{\psi\psi}} \,d\,\phi \right)  \nonumber \\ [3mm]
\omega\,^{\hat{r}} & =&  (\,g_{rr}\,)\,^{1/2}\,d\,r \nonumber\\[3mm]
\omega\,^{\hat{\theta}} & =&  (\,g_{\theta\theta}\,)\,^{1/2}\,d\,\theta \,\,,
\label{frame}
\end{eqnarray}
while the dual basis is given by
\begin{eqnarray}
e\,_{\hat{t}} & = & \left| g_{tt}- \frac{{g_{t\psi}}^2}{g_{\psi\psi}}
-\Omega_{(a)}^2\,\,\frac{g_{\phi\phi}\,g_{\psi\psi}-{g_{\phi\psi}}^2}
{g_{\psi\psi}} \right|^{-1/2}\,
\left(\frac{\partial}{\partial t} + \Omega_{(a)}\,
\frac{\partial}{\partial \phi} -
\frac{g_{t\psi}+ \Omega_{(a)}\,g_{\phi\psi}}{g_{\psi\psi}}\,\,
\frac{\partial}{\partial \psi}\right) \nonumber\\[3mm]
e\,_{\hat{\phi}} & =& \left(\frac{g_{\phi\phi}\,g_{\psi\psi}
-{g_{\phi\psi}}^2} {g_{\psi\psi}} \right)^{\,-1/2} \,
\left(\frac{\partial}{\partial \phi} - \frac{g_{\phi\psi}}{g_{\psi\psi}}\,\,
\frac{\partial}{\partial \psi} \right)
\nonumber\\[3mm]
e\,_{\hat{\psi}} & =& \frac{1}{(\,g_{\psi\psi}\,)\,^{1/2}}\,
\frac{\partial}{\partial \psi} \nonumber \\ [3mm]
e\,_{\hat{r}} & =&  \frac{1}{(\,g_{rr}\,)\,^{1/2}}
\,\frac{\partial}{\partial r} \nonumber\\[3mm]
e\,_{\hat{\theta}} & =&  \frac{1}{(\,g_{\theta\theta}\,)\,^{1/2}}
\,\frac{\partial}{\partial \theta}
\label{dualframe}
\end{eqnarray}

We note that the corresponding basis $1$-forms and their duals
associated with an orthonormal frame of a locally nonrotating
observer moving in the $\,\psi\,$ direction are obtained
from the expressions given above simply by the transformation
$\,\phi \leftrightarrow \psi\,$.

\begin{center}
\section{Uniform Magnetic Field in the Background of a Five Dimensional
Black Hole}
\end{center}

\subsection{Uniform magnetic field in  a five dimensional flat
spacetime}

In four dimensional gravity the behaviour of  electromagnetic fields
in the background of a rotating black hole described by the Kerr metric
has been investigated by many authors \cite{wald, klk, ernst, ew}
(see also Ref.\cite{ag} for a review). In particular, Wald has
proposed the most elegant way of desribing the behaviour of
electromagnetic fields
near a rotating black hole which is placed in an originally
uniform magnetic field aligned with the axis of symmetry of
the black hole \cite{wald}. The Wald approach stems from the well known
fact \cite{papa} that a Killing vector in a vacuum spacetime serves as
a vector potential for a Maxwell test field in that spacetime's background.
Therefore one can construct a solution for the Maxwell test field
in the background of a vacuum spacetime simply by using the
isometries of this spacetime.

We shall apply this approach to examine the behaviour of a Maxwell test
field around a five dimensional black hole described by the
metric (\ref{metric}) when the black hole is immersed in an asymptotically
uniform magnetic field. We recall that in general,
the electromagnetic field in five dimensions
can be described by an electric $1$-form field and magnetic $2$-form field.
In other words, four of the total ten independent components of the
electromagnetic field tensor $\,F_{\mu\nu}\,$ describe
the electric field, while the remaining six components correspond
to the magnetic field. We shall consider the case of a magnetic
field configuration
which is stationary and uniform at infinity
and possesses bi-azimuthal symmetry as well.
Then it is clear that the corresponding electromagnetic field tensor must
share all isometries of the black hole's spacetime. On these grounds,
it is also clear that there must exist only two nonvanishing
components of the field tensor that describe a uniform
magnetic field in our case. That is, we have
\begin{eqnarray}
B&=&F_{xy}\,\,,~~~~~~~~  H=F_{zw}\,,
\label{field}
\end{eqnarray}
where we have used the notations $\,B\,$ and $\,H\,$ for the magnetic
field strengths associated with the (x-y) and (z-w)
$2$-planes, respectively. We note that these quantites are
reminiscent of the two independent angular momenta
given in (\ref{momenta}), which
are, in turn, the nonvanishing components of the underlying
angular momentum $2$-form. In this respect, the magnetic field
components in our model are aligned with the corresponding
angular momenta.

In the following we shall need the expression
for the components of the electromagnetic $2$-form field
$\,F\,$ written down in bi-polar coordinates.
By making use of the transformations (\ref{cartesian}) we obtain, instead of
(\ref{field}), an expression of the form
\begin{eqnarray}
F&=& B \,r\sin\theta \left(\sin\theta \,d\,r \wedge d\,\phi
+ r \cos\theta  \,d\,\theta \wedge d\,\phi \right)
\nonumber \\ [2mm] & &
+ \,H\,r \cos\theta \left(\cos\theta \,d\,r \wedge d\,\psi
- r\,\sin\theta \,d\,\theta \wedge d\,\psi \right)\,\,,
\label{f2form}
\end{eqnarray}
which describes a uniform magnetic field of a configuration
with bi-azimuthal symmetry in a flat five dimensional spacetime.

\subsection{$5$-Vector Potential}

It is remarkable that using only the isometries of the Myers-Perry
spacetime described by the temporal Killing vector $\,{\bf \xi}_{(t)} \,$
and two azimuthal Killing vectors $\,{\bf \xi}_{(\phi)}\,$ and
$\,{\bf \xi}_{(\psi)}\,$
one can construct $5$-vector potential for the Maxwell test field
in this spacetime. Indeed, the homogeneous Maxwell equations,
in the Lorentz gauge
\begin{equation}
A^{\mu}_{\;;\;\mu} =0\,,
\label{lorentz}
\end{equation}
have the form
\begin{equation}
{A^{\mu\,;\,\nu}}_{\,;\,\nu}- {R^{\mu}}_{\nu}\,A^\nu=0\,\,.
\label{max}
\end{equation}
On the other hand, any Killing vector $\, {\bf \xi}\,$ satisfies
the equation
\begin{equation}
{\xi^{\mu\,;\,\nu}}_{\,;\,\nu} + {R^{\mu}}_{\nu}\,\xi^\nu=0\,\,.
\label{kileq}
\end{equation}
Comparing the two equations (\ref{max}) and (\ref{kileq}) one sees that
they are the same in a vacuum spacetime $\,({R^{\mu}}_{\nu}=0)\,$.
Thus, one can use a Killing vector as a vector potential for a test
Maxwell field \cite{papa}.   Following this fact, we shall seek for a
$5$-vector potential of the form
\begin{equation}
A^{\mu} = \alpha \,\, \xi^{\mu}_{(t)}+ \beta \,\, \xi^{\mu}_{(\phi)}
+\gamma \, \xi^{\mu}_{(\psi)} \,\,,
\label{potform}
\end{equation}
where $\,\alpha\,$, $\,\beta\,$  and  $\,\gamma\,$ are arbitrary parameters.
Let us emphasize that in a general case this vector potential
will also describe a test electric field so that the black hole is
charged. We assume that this electric charge is small enough,
$\,Q \ll M\,$, so that the spacetime can still be adequately described by
the unperturbed metric (\ref{metric}).

To determine the unknown parameters in (\ref{potform}) first
we calculate the corresponding electromagnetic $2$-form field $\,F\,$
in the metric (\ref{metric}). We obtain
\begin{eqnarray}
F&=&\,\frac{2\,m\,r}{\Sigma\,^2}\,{\cal{A}}\,d\,t \wedge d\,r
 \nonumber\\[3mm] & &
+ \frac{m\,\sin2\theta}{\Sigma\,^2}\,\left[\alpha\,(b^2-a^2)
+\beta\,a\,(r^2+a^2) - \gamma\,b\,(r^2+b^2)\right]\,d\,t \wedge d\,\theta
 \nonumber\\ [3mm]   & &
+ 2\,r\,\sin^2\theta \left(\beta +
\frac{a\,m}{\Sigma\,^2}\,\cal{A}\right)\,d\,r \wedge d\,\phi
 + 2\,r\,\cos^2\theta \left(\gamma +
\frac{b\,m}{\Sigma\,^2}\,\cal{A}\right)\,d\,r \wedge d\,\psi
 \nonumber\\  [3mm]        & &
+ \sin2\theta\,\left[\beta\,(r^2+a^2) +
\frac{a\,m}{\Sigma\,^2}\,\left({\cal{B}} - \alpha\,(r^2+a^2) - \gamma\,b\,
(r^2+b^2) \right) \right] \,d\,\theta \wedge d\,\phi
 \nonumber\\       [3mm]       & &
- \sin2\theta\,\left[
\gamma\,(r^2+b^2) +
\frac{b\,m}{\Sigma\,^2}\,\left( {\cal{B}} - \alpha\,(r^2+b^2) - \beta\,a\,
(r^2+a^2) \right) \right] \,d\,\theta \wedge d\,\psi  \,\,,
\label{fmu}
\end{eqnarray}
where for the sake of brevity we have used the notations
\begin{equation}
{\cal{A}}=\alpha - \beta\,a \,\sin^2\theta - \gamma\,b\,\cos^2\theta\,\,
\end{equation}
and
\begin{equation}
{\cal{B}}=  \beta\,a \,(r^2+a^2+\Sigma)\,\sin^2\theta + \gamma\,b\,
(r^2+b^2+\Sigma)\,\cos^2\theta \,\,.
\end{equation}
In the asymptotic region  $\,r \rightarrow \infty\,$\,, this expression
takes the form
\begin{eqnarray}
F&=& 2\,\beta \,r\sin\theta \left(\sin\theta \,d\,r \wedge d\,\phi
+ r \cos\theta  \,d\,\theta \wedge d\,\phi \right)
\nonumber \\ [2mm] & &
+ \,2\, \gamma\,r \cos\theta \left(\cos\theta \,d\,r \wedge d\,\psi
- r\,\sin\theta \,d\,\theta \wedge d\,\psi \right)\,\,
\label{f2form11}
\end{eqnarray}
From a comparison of equations (\ref{f2form}) and (\ref{f2form11})
it follows that $\,\beta=B/2\,,\,\, \gamma=H/2\,\,$. The remaining parameter
$\,\alpha\,$ can be determined from examining the
integrals (\ref{mass}) and
(\ref{momenta}) along with the integral for the electric charge of
the black hole
\begin{equation}
Q=\frac{1}{4\,\pi^2}\,\oint \,F^{\mu\,\nu}\,
d\,^3 \Sigma_{\mu\;\nu} \,\,.
\label{charge}
\end{equation}
Having done this, we obtain 
\begin{equation}
\alpha=-\frac{Q}{2\,m} + \frac{a\,B}{2}  + \frac{b\,H}{2}\,\,.
\label{alpha}
\end{equation}
Finally, the $5$-vector potential for the electromagnetic  field
around a rotating and weakly charged black hole in a uniform magnetic field
can be written in the form
\begin{equation}
A^{\mu} = - \frac{Q}{2\,m} \,\,\xi^{\mu}_{(t)}
+ \frac{B}{2} \left(\xi^{\mu}_{(\phi)} + a\,\xi^{\mu}_{(t)}\right)
+ \frac{H}{2} \left(\xi^{\mu}_{(\psi)} + b\,\xi^{\mu}_{(t)}\right)\,\,.
\label{fpotform}
\end{equation}
From this expression, it follows that the  $5$-vector potential
in the background of the Myers-Perry spacetime consists of a superposition
of the Coulomb-type part and the asymptotically uniform magnetic field.
It is important to note that the Coulomb-type part, which is generated
by the temporal Killing vector, does not vanish even when the electric
charge of the black hole is zero $\,(Q=0)\,$. There are two independent
contributions to the Coulomb field of the black hole.
The physical reason underlying this phenomenon is
that a rotation of a five dimensional
black hole in a uniform magnetic field produces an
inductive electric field associated with the two independent
$\,\phi\,$ and $\,\psi\,$ $2$-planes of rotation of the black hole
as well as two independent components of the electromagnetic field tensor.
It is seen that in our model
the black hole acts like a "dynamo" that causes an electrostatic
potential difference between the event horizon of the black hole
and an infinitely distant surface. Following the approach in
four dimensional case \cite{carter1} we shall define the electrostatic
potential of the event horizon with respect to the Killing vector
(\ref{khorizon}) as follows
\begin{eqnarray}
\Phi_h & = & A\cdot \chi   =
A_0 + \Omega_{(a)\,h}\,A_{\phi} +\Omega_{(b)\,h}\,A_{\psi} \,\,.
\label{hpot}
\end{eqnarray}
Then, for the electrostatic potential difference between
the event horizon and an infinitely distant surface, we find
\begin{eqnarray}
\triangle \Phi & = &
\Phi_h - \Phi_{\infty}= \,\frac{Q - a\,m B - b\,m H }{2\,m} \,\,.
\label{potdif}
\end{eqnarray}
This potential difference is exactly of the same form as if it
were be produced by the electric charge
\begin{equation}
{\tilde Q} = Q - a\,m B - b\,m H\,\,.
\label{indcharge}
\end{equation}

It is obvious that this charge $\tilde{Q}$ will be quickly
neutralized (the potential difference vanishes) due to a selective
accretion process of charged particles provided that the black hole
is surrounded by an ionized medium\footnote{In the brane world
scenario, when charged particles can live only on the brane, this
conclusion also remains true. In the presence of a plasma of charged
particles located on the brane there will be a selective
accretion process reducing the charge $\tilde{Q}$.}
As a result of this the black hole will acquire the physical electric charge
\begin{eqnarray}
Q &=& a\,m B + b\,m H = j_{(a)} B + j_{(b)} H \,\,.
\label{indcharge1}
\end{eqnarray}
We note that in the degenerate case when $\,a=b \,$ and $\,B=H \,$ the above
expression goes over into the form
\begin{equation}
Q = 2\,a\,m B,
\label{indcharge2}
\end{equation}
which is reminiscent of its counterpart for a Kerr black hole \cite{wald}.

Accordingly, in terms of the total mass and angular momenta
of the black hole the equation (\ref{indcharge1})
can be written as
\begin{eqnarray}
Q&=&\frac{8}{3\,\pi} \left(a\,M B + b\,m H \right)=
\frac{4}{\pi} \left(J_{(a)} B +J_{(b)}\, H \right)
\label{indcharge3}
\end{eqnarray}
Thus, a five dimensional black hole rotating in a uniform magnetic field
of configuration with bi-azimuthal symmetry will charged up to the
value given by equation (\ref{indcharge3}).

To conclude this section we shall calculate the magnetic flux crossing
a portion $\,\Sigma \,$ of the black hole event horizon.
This flux is governed by the line integral on the horizon
\begin{equation}
{\cal F} = \int_{\partial \Sigma} A\,\,,
\label{flux}
\end{equation}
where the potential $1$-form $\,A\,$ is determined through
(\ref{fpotform}) with $\,Q=0\,$ and $\partial \Sigma$ is the boundary of
$\,\Sigma \,$. For our purpose, it is convenient to
rewrite the potential in terms of
the Killing vector $\,\chi\,$  defined in Eq. (\ref{khorizon}). We
obtain
\begin{eqnarray}
A &=& \frac{1}{2}\,\left(a\,B + b\,H \right)\,\chi +\tilde{A}\,,
\nonumber\\  [3mm]
\tilde{A} &=&
 \frac{B}{2}\left({\bf \xi}_{(\phi)} - a\,
\Omega_{(a)\,h}\,{\bf \xi}_{(\phi)} -
a\,\Omega_{(b)\,h} \,{\bf \xi}_{(\psi)}\right)
+ \frac{H}{2}\left({\bf \xi}_{(\psi)} - b\,
\Omega_{(b)\,h}\,{\bf \xi}_{(\psi)} -
b\,\Omega_{(a)\,h} \,{\bf \xi}_{(\phi)}\right)\,\,,
\label{flux1}
\end{eqnarray}
where we have used the same notation for Killing $1$-form fields.
The first term in this expression is proportional to  $\,\chi\,$ and
therefore its contribution to the flux ${\cal F}$ at the horizon vanishes.
For the rest part, taking into account
(\ref{horizon})  we obtain
\begin{equation}
\tilde{A} = \frac{1}{2}\,\left(\frac{B\,r_{h}^2 - H\,a\,b}{r_{h}^2+a^2}\,\,
{\bf \xi}_{(\phi)}
+\frac{H\,r_{h}^2 - B\,a\,b}{r_{h}^2+b^2}\,\,{\bf \xi}_{(\psi)}
\right) \,\,.
\label{flux2}
\end{equation}
This expression, and hence, the flux (\ref{flux})
vanishes precisely at the extremal horizon of the black
hole $\, (r_{h}^2 =a \,b)\,$, provided that $\,B=H\,$. Thus, when the
magnetic field strengths associated with the
(x-y) and (z-w) $2$-planes of rotation are equal in
magnitude, the magnetic flux is expelled from a five dimensional black hole
as the extremality in its rotation is approached. In this case, a portion
of the black hole horizon, like its four dimensional counterpart,
acts as the surface of a perfectly diamagnetic object \cite{ceg}.

\section{Magnetic Dipole Moments}

We now turn to the consideration of a five dimensional weakly
charged rotating black hole. It is clear that the rotation of
such a black hole must produce a dipole
type magnetic field around itself. Since the charged black hole
is characterized by two independent rotation parameters,
accordingly, one may expect that it
will acquire two independent magnetic
dipole moments as well. We shall determine the value of these
magnetic moments.

We begin with the expression for the electromagnetic $2$-form field

\begin{eqnarray}
F&=&\,\frac{Q}{\Sigma\,^2}\,\left(r\,d\,r \wedge d\,t
+(b^2-a^2)\,\sin\theta \cos\theta \,d\,\theta \wedge d\,t \right)
 \nonumber\\[3mm] & &
- \frac{Q\,a\,\sin\theta}{\Sigma\,^2}\,\left(r\,\sin\theta\,
d\,r\wedge d\,\phi
- (r^2+a^2)\,\cos\theta\,d\,\theta  \wedge d\,\phi \right)
 \nonumber\\ [3mm]   & &
- \frac{Q\,b\,\cos\theta}{\Sigma\,^2}\,\left(r\,\cos\theta\,
d\,r\wedge d\,\psi
+ (r^2+b^2)\,\sin\theta\,d\,\theta  \wedge d\,\psi \right) \,\,,
\label{eft}
\end{eqnarray}
which is obtained from (\ref{fmu})  with $\,B=0\,,\,\,\,
H=0\,\,$. The associated potential $1$-form can be written as
\begin{equation}
A = -\frac{Q}{2\,\Sigma}\, \left(d\,t - a\,\sin^2\theta\,d\,\phi
- b\,\cos^2\theta\,d\,\psi\right)\,\,.
\label{potform1}
\end{equation}
In obtaining this expression we have gauged the potential
(\ref{fpotform}) according to the transformation
\begin{equation}
A = {\hat{A}} - \frac{Q}{2\,m} \,d\,t\,\,
\label{cgauge}
\end{equation}
to provide its vanishing behaviour at infinity. We shall also
need the contravariant components of the electromagnetic
field tensor which  are given by
\begin{eqnarray}
F^{tr}&=&\frac{Q\,(r^2+a^2)(r^2+b^2)}{r\,\Sigma\,^3}\,\,,
~~~~~
F^{t\theta}= \frac{Q\,(b^2-a^2)\,\sin2\theta}{2\,\Sigma\,^3}\,\,,
\nonumber \\[3mm]
F^{r\phi}& =& - \frac{Q\,a\,(r^2+b^2)}{r\,\Sigma\,^3}\,\,,
~~~~~~~~~~~~~~~
F^{r\psi} = - \frac{Q\,b\,(r^2+a^2)}{r\,\Sigma\,^3}\,\,,
 \nonumber \\[3mm]
F^{\theta\phi}& =&\frac{Q\,a\,\cot\theta}{\Sigma\,^3}\,\,,
~~~~~~~~~~~~~~~~~
F^{\theta\psi} = - \frac{Q\,b\,\tan\theta}{\Sigma\,^3}\,\,.
\label{emtcontra}
\end{eqnarray}

Next, we shall define the electric field, as well as the dipole magnetic
field in the asymptotic rest frame of the black hole. We start with
the electric $1$-form field $\,{\hat E}\,$ which,
in the spacetime of dimensions $\,D\,$, can be defined as follows
\begin{eqnarray}
{\hat E} &=& - i_{{\xi}_{(t)}} F = (-1)^D\,
^{\star}\left(\xi_{(t)}\wedge ^{\star}F\right)\,\,,
\label{e1form}
\end{eqnarray}
where $\,\xi_{(t)} = {\xi}_{(t) \mu}\,d\,x^\mu\,$ is
the timelike Killing $1$-form field and the $\,\star\,$
operator denotes the Hodge dual. Substituting (\ref{eft})
in equation (\ref{e1form}) we obtain the following expression
for the electric $1$-form field in the metric (\ref{metric})
\begin{equation}
{\hat E} = \frac{Q}{\Sigma\,^2} \left(r\,d\,r
+ (b^2-a^2) \,\sin\theta \cos\theta\, d\,\theta \right)\,\,.
\label{e1}
\end{equation}
The orthonormal components of the electric field in the asymptotic
rest frame of the black hole are obtained by projecting (\ref{e1})
on the basis (\ref{dualframe}). We have
\begin{eqnarray}
E\,_{\hat r} & = & F_{\hat{r}\hat{t}}= \frac{Q}{r^3} +
\mathcal{O}\left(\frac{1}{r^5}\right)\,\,,
\nonumber \\[3mm]
E\,_{\hat \theta} & = & F_{\hat{\theta} \hat{t}}=
\mathcal{O}\left(\frac{1}{r^5}\right)\,\,,
\nonumber \\ [3mm]
E\,_{\hat \phi}& =& F_{\hat{\phi}\, \hat{t}}=0
\label{efiled}
\end{eqnarray}
We note that the dominant component of the electric field is purely
radial and the associated Gaussian flux of this radial field
gives the correct value for the electric charge of the black hole.

The dipole magnetic field of the black hole is described by
the magnetic $2$-form defined as
\begin{eqnarray}
{\hat B} &=& -i_{\xi_{(t)}}\, ^{\star}F =
^{\star}\left(\xi_{(t)}\wedge F\right)\,\,,
\label{m2form}
\end{eqnarray}
This can also be rewritten in the alternative form
\begin{eqnarray}
{\hat B} = \frac{1}{4}\,\sqrt{-g}\,\epsilon_{\mu\,\nu\,\alpha\,\beta\,
\gamma} \,\xi_{(t)}^\mu \,F^{\nu\alpha}\,
d\,x^{\beta} \wedge d\,x^{\gamma}\,\,,
\label{m2form11}
\end{eqnarray}
Substituting into this expression the contravariant components of the
electromagnetic field tensor (\ref{emtcontra}) we obtain
\begin{eqnarray}
{\hat B} &=& \frac{Q\,b\,\sin\theta}{\Sigma\,^2}\,\left
(r \sin\theta\, d\,r \wedge d\,\phi
- (r^2+a^2)\,\cos\theta\, d\,\theta \wedge d\,\phi\right)
\nonumber\\[3mm] & &
+ \frac{Q\,a\,\cos\theta}{\Sigma\,^2}\,\cos\theta \left(r \cos\theta \,
d\,r \wedge d\,\psi
+ (r^2+b^2)\,\sin\theta\, d\,\theta \wedge d\,\psi\right)\,\,
\label{m2form1}
\end{eqnarray}
which in the asymptotic rest frame of the black hole has the
the following orthonormal components
\begin{eqnarray}
B_{{\hat r}{\hat\psi}}  & = & F_{\hat{\theta}\hat{\phi}}=
\frac{F_{\theta\phi}}{r^2\,\sin\theta}=
\frac{Q\,a}{r^4}\,
\cos\theta +
\mathcal{O}\left(\frac{1}{r^6}\right)\,\,,
\nonumber \\[3mm]
B_{{\hat\theta}{\hat\psi}}  & = & F_{\hat{\phi}\hat{r}}=
\frac{F_{\phi r}}{r\,\sin\theta}= \frac{Q\,a}{r^4}\,
\sin\theta +
\mathcal{O}\left(\frac{1}{r^6}\right)\,\,,
\nonumber \\[3mm]
B_{{\hat\phi}{\hat r}} & = & F_{\hat{\theta}\hat{\psi}}=
\frac{F_{\theta \psi}}{r^2\,\cos\theta}=
 -\frac{Q\,b}{r^4}\,
\sin\theta +
\mathcal{O}\left(\frac{1}{r^6}\right)\,\,,
\nonumber \\[3mm]
B_{{\hat\theta}{\hat\phi}}  & = & F_{\hat{r}\hat{\psi}}=
\frac{F_{r \psi}}{r\,\cos\theta}=
-\frac{Q\,b}{r^4}\,
\cos\theta +
\mathcal{O}\left(\frac{1}{r^6}\right)\,\,,
\label{mfiled1}
\end{eqnarray}
with all others vanishing. The above expressions describe the dipole magnetic
fileld created by a five dimensional weakly charged rotating black hole.
We see that far from the black hole the dominating behaviour of
the magnetic field is determined only by the two independent quantities
\begin{eqnarray}
\mu_{(a)} & = & Q\,a\,\,,~~~~~~  \mu_{(b)}  =  Q\,b\,\,,~~~~~~
\label{dipole}
\end{eqnarray}
which can be thought of as the magnetic dipole moments of the black hole.
We conclude that a weakly charged rotating black hole in
five dimensions possesses two independent magnetic moments
specified only in terms of
the electric charge of the black hole and its two rotation parameters.

\section{Gyromagnetic Ratio}

It is now natural to address the gyromagnetic ratio of the five dimensional
weakly charged rotating black hole we discussed above. We recall
that one of the remarkable facts about a charged rotating black hole
of four dimensional general relativity is that it can be assigned
a gyromagnetic ratio $\,g=2\,$\, just like  the electron
in Dirac theory \cite{carter2}. The parameter $\,g\,$ is defined as
a constant of proportionality in the equation
\begin{equation}
\mu= g\, \frac{Q\,J}{2 \,M}\,,
\label{g1}
\end{equation}
where $\,M\,$ is the mass, $\,J\,$ is the angular momentum and
$\,Q\,$ is the electric charge of the four dimensional black hole.

Turning now to the case of a weakly charged black hole in five dimensions
and comparing equations (\ref{momenta}) and (\ref{dipole}) we
see that the coupling of rotation parameters
of the black hole to its mass parameter to give the
specific angular momenta looks exactly the same as
their coupling to the electric charge to define the
magnetic dipole moments. Thus, we may
write the reminiscent of equation (\ref{g1}) in five dimensions as follows
\begin{eqnarray}
\mu_{(i)}&=&\frac{Q\,j_{(i)}}{m} = 3\,\frac{Q\,J_{(i)}}{2 \,M}\,,
\label{g2}
\end{eqnarray}
where we have used the relations (\ref{physical}) and the subscript index
$\,i\,$ refers to either the parameter $\,a\,$\, or, $\,b\,$.
From  a comparison of this equation with the classical relation
(\ref{g1}) it becomes apparent that a five dimensional weakly
charged rotating black hole can be assigned a gyromagnetic ratio $\,g=3\,$.

Next, following the basic arguments of \cite{wald} we shall prove
the value $\,g=3\,$. For this purpose, we shall define the twist
\cite{heusler} of a timelike Killing $1$-form field
$\,\xi_{(t)}\,$, which in five dimensions is the $2$-form field
given by
\begin{equation}
\Omega  =  \frac{1}{2}\,^{\star}\left({\hat\xi}_{(t)}
\wedge d\,{\hat\xi}_{(t)}\right)\,,
\label{twist2}
\end{equation}
Physically this quantity measures the failure of
the timelike Killing  $1$-form field to be hypersurface orthogonal.
Evaluating this quantity in the metric (\ref{metric})
we obtain
\begin{eqnarray}
\Omega&=& - \frac{b\,m\,r}{\Sigma\,^2}\,\sin^2\theta\,d\,r \wedge d\,\phi
+ \frac{b\,m\,(r^2+a^2)}{\Sigma\,^2}\,\sin\theta \cos\theta\,
d\,\theta \wedge d\,\phi
- \frac{a\,m\,r}{\Sigma\,^2}\,\cos^2\theta\,d\,r \wedge d\,\psi
\nonumber\\[3mm] & &
- \,\frac{a\,m\,(r^2+b^2)}{\Sigma\,^2}\,\sin\theta \cos\theta\,
d\,\theta \wedge d\,\psi \,\,,
\label{mptwist2}
\end{eqnarray}
which implies the existence of the twist potental $1$-form
\begin{equation}
\omega =  \frac{m}{2\,\Sigma}\,\left(b\,\sin^2\theta\, d\,\phi+
a\,\cos^2\theta\, d\,\psi \right)
\label{twistpot}
\end{equation}
The components of this quantity in the asymptotic rest frame of
the black hole show that the failure of the
timelike Killing vector to be hypersurface orthogonal
is completely determined by the specific angular momenta
$\,j_{(a)}=a\, m\,$ and $\,j_{(b)}= b\, m\,$ of the black hole.

On the other hand, the magnetic $2$-form field (\ref{m2form1})
implies the magnetic potential $1$-form determined through
the equation
\begin{equation}
B = -d\,\varphi \,\,,
\label{mstatpot}
\end{equation}
where
\begin{equation}
\varphi =  \frac{Q}{2\,\Sigma}\,\left(b\,\sin^2\theta\, d\,\phi+
a\,\cos^2\theta\, d\,\psi \right)\,\,.
\label{mstatpot1}
\end{equation}
From this expression it follows that in the asymptotic rest frame
of the black hole the magnetic potential $1$-form determines
the two magnetic dipole moments, just as the twist $1$-form (\ref{twistpot})
determines the two specific angular momenta of the black hole.
From equations (\ref{twistpot}) and (\ref{mstatpot1}) we read off
the relation
\begin{equation}
\varphi =  \frac{Q}{m}\,\omega \,\,,
\label{varphi}
\end{equation}
which, obviously, can also be rewritten in the form of (\ref{g2}).
This proves that a five dimensional Myers-Perry
black hole endowed with a small enough electric charge must have
a gyromagnetic ratio of value $\,g=3\,$.
The same value of gyromagnetic ratio has been found for a supersymmetric
rotating black hole \cite{herdeiro} described by the
Breckenridge-Myers-Peet-Vafa (BMPV) five dimensional solution \cite{bmpv}.

\section{Conclusions}

We have discussed the special properties of a five dimensional
rotating black hole described by the Myers-Perry metric in the presence
of an originally uniform magnetic field. The configuration of the
magnetic field is supposed to have bi-azimuthal symmetry,
just like the black hole spacetime itself. In this case
the magnetic field has only two nonvanishing components aligned with the
two angular momenta of the black hole. We have also allowed
the black hole to have an electric charge small enough
that the spacetime can still be described by the Myers-Perry solution.

We have constructed the 5-vector potential describing
the test Maxwell field in the Myers-Perry spacetime using
the Killing isometries of this spacetime. The intriguing feature
of this model is the appearance of non-trivial gravitomagnetic
phenomena; a rotation of a five dimensional black hole in a
uniform magnetic field of given configuration produces an
inductive electrostatic potential difference between the event horizon
and an infinitely distant surface. This potential difference comes from
the superposition of two independent Coulomb fields
arising due to rotations in two distinct $2$-planes and two nonvanishing
components of the magnetic field. Of course, in the case of an
ionized medium surrounding the black hole, the potential difference
will be quickly neutralized by a selective accretion process, thereby
providing a mechanism for charging up the black hole.

We have also described a dipole magnetic field around a weakly charged
rotating black hole in five dimensions and as expected,
it turned out that the black hole possesses two independent
magnetic dipole moments determined only by
its electric charge, mass and angular momentum parameters.
In many aspects gravitomagnetic phenomena described are qualitatively
closly reminiscent of their counterparts for a four dimensional
Kerr black hole immersed in a uniform magnetic field. However, there
also exist some essential differences. In particular, we have shown
that the gyromagnetic ratio for a five dimensional weakly charged
Myers-Perry black hole is $g=3$.

In four dimensional gravity there exist stable circular orbits
in the equatorial plane of a Kerr black hole. Furthermore, the
presence of a uniform magnetic field around the  Kerr
black hole has its greatest effect in enlarging the region of stability
of the circular orbits toward the horizon \cite{an}.
However, there are no stable
circular orbits around a five dimensional rotating black hole, at least
in the equatorial planes \cite{fs1}. Therefore, it would be interesting
to use the results of this paper to study the effect of an external magnetic
field on the stability of circular motion around a five dimensional
Myers-Perry metric.

\begin{acknowledgments}
A. N. thanks Gary Gibbons and Bahram Mashhoon for useful discussions
on gyromagnetic ratio at an early stage of this work. V.F. thanks
Feza G\"ursey Institute for kind hospitality during the Research
Semester Gravitation and Cosmology, June 2003. V.F is also grateful to
the Natural Sciences and Engineering Research Council of Canada and the
Killam Trust for their financial support.

\end{acknowledgments}

\end{document}